\documentclass[prc,onecolumn,superscriptaddress,showpacs,psfig,showkeys]{revtex4}
\usepackage{amsmath}
\usepackage{amssymb}
\usepackage{amsfonts}
\usepackage{graphicx}
\usepackage{dcolumn}
\usepackage{epstopdf}
\usepackage{color}
\usepackage[utf8]{inputenc}
\usepackage{amsthm}
\usepackage{fontenc}
\usepackage{bm}
\RequirePackage{slashed}\usepackage{cancel}
\usepackage[normalem]{ulem}

\usepackage{array,booktabs}

\newcolumntype{C}{>{$\displaystyle}c<{$}}

\begin{document}

\pacs{13.60.Hb,12.39.Ki,12.38.Bx}
\keywords{Deeply Virtual Compton scattering, nuclear parton structure}

\title{
Coherent deeply virtual Compton scattering off $^4$He
}

\author{Sara Fucini}
\affiliation{ Dipartimento di Fisica e Geologia,
Universit\`a degli Studi di Perugia and Istituto Nazionale di Fisica Nucleare,
Sezione di Perugia, via A. Pascoli, I - 06123 Perugia, Italy}
\author{Sergio Scopetta}
\affiliation{ Dipartimento di Fisica e Geologia,
Universit\`a degli Studi di Perugia and Istituto Nazionale di Fisica Nucleare,
Sezione di Perugia, via A. Pascoli, I - 06123 Perugia, Italy}
\author{Michele Viviani}
\affiliation{
INFN-Pisa, 56127 Pisa, Italy} 
\date{\today}

\begin{abstract}
\vspace{0.5cm}
Coherent Deeply virtual Compton scattering off the $^4$He nucleus 
is studied in impulse
approximation.
A convolution formula for the nuclear Generalized Parton Distribution
(GPD) is derived in terms of the $^4$He non-diagonal spectral function
and of the GPD of the struck nucleon.
A model of the nuclear non diagonal spectral function, 
based on the momentum distribution
corresponding to the AV18 nucleon-nucleon interaction, is used in the actual
calculation. Typical impulse approximation results
are reproduced, in proper limits, for the nuclear form factor
and for nuclear parton distributions.
The nuclear generalized parton distribution and the
Compton form factor are evaluated using, as nucleonic ingredient,
a well known generalized parton distribution model. 
An overall very good agreement is found with the data recently published
by the EG6
experiment at the Jefferson Laboratory (JLab).
More refined nuclear calculations are addressed and will be necessary for 
the expected improved accuracy of the next generation of experiments 
at JLab, with the 12 GeV electron beam and high luminosity. 

\end{abstract}

\maketitle

\section{Introduction}
Nuclear Generalized Parton Distributions (GPDs), 
measured in hard-exclusive
electroproduction processes in nuclei,
can provide a wealth of novel information
(for a recent report, see, i.e. 
\cite{Dupre:2015jha}),
such as a signature of the presence of non-nucleonic degrees of freedom
\cite{Berger:2001zb}
or a nuclear tomography, i.e., the distribution
of partons with a given longitudinal momentum
in the nuclear transverse plane.
Nuclear GPDs
can be therefore very important
for a fully quantitative explanation of the so-called EMC effect
\cite{Aubert:1983xm},
i.e., the nuclear modifications of the parton structure of bound
nucleons (see Ref. \cite{Hen:2013oha} for a recent report).

Several processes can be described in terms of GPDs. Among them,
the one of interest here is coherent Deeply Virtual Compton Scattering
(DVCS), i.e. deep exclusive photon electroproduction off a nuclear
target $A$, the hard fully exclusive reaction $A(e,e'\gamma)A$, which
could give access to the quark tomography of the nucleus as a whole.
The experimental study of this process requires the very difficult
coincidence detection 
of fast photons and electrons
together with slow, intact recoiling nuclei. 
For this reason, in the first measurement of nuclear DVCS at
HERMES \cite{Airapetian:2009cga}, a clear separation 
was not achieved between the
coherent process and the so-called incoherent one,
i.e. the process $A(e,e'\gamma N)X$, which allows 
the tomography of the bound nucleon. The latter, compared with
that of the free nucleon, could provide a pictorial view of the realization 
of the EMC effect. 

Much theoretical work has been performed
to study nuclear GPDs (see Ref. \cite{Dupre:2015jha} for a review
of results). We remind that,
measuring GPDs through DVCS,
it has been suggested to study
the distribution of nuclear forces in nuclei
\cite{Polyakov:2002yz,Kim:2012ts,Jung:2014jja}  
and the
modifications of the bound nucleon structure 
\cite{Freund:2003ix,Freund:2003wm,
Liuti:2004hd,Guzey:2005ba,
Liuti:2005qj,Goeke:2009tu,Guzey:2008th,
Guzey:2008fe,Guzey:2009pv}.
The general formalism of DVCS on nuclear targets of any spin
has been developed initially in Ref. \cite{Kirchner:2003wt}.

In these studies, a special role is played
by few nucleon systems,
such as $^2$H, $^3$He, $^4$He.
As a matter of fact, although challenging, for these targets
a realistic evaluation of 
conventional nuclear effects is possible.
This would allow to distinguish these effects
from exotic ones, which could be responsible of
the observed EMC behaviour. Without realistic benchmark calculations,  
the interpretation of the collected data will be hardly conclusive. 
In these sense, the use of heavier targets,
due to the difficulty of the corresponding
realistic many-body calculations, has a weaker priority.
The $^2$H nucleus
is very interesting, for the extraction of the neutron information
and for its rich spin structure
\cite{Berger:2001zb,Cano:2003ju,Taneja:2011sy}.
In between $^2$H and $^4$He, $^3$He could allow
to study the $A$ dependence of nuclear effects and
it could give an easy access to neutron polarization properties,
due to its specific spin structure. Besides, being not isoscalar, 
flavor dependence of nuclear effects could be studied, 
in particular if parallel measurements 
on $^3$H targets were possible.
A complete impulse approximation (IA)
analysis, using the Argonne 18 (Av18)
nucleon-nucleon potential
\cite{Wiringa:1994wb}, is available and
nuclear effects on GPDs are found to be sensitive to details of 
the used nucleon-nucleon interaction
\cite{Scopetta:2004kj,
Scopetta:2009sn,
Rinaldi:2012pj,Rinaldi:2012ft,
Rinaldi:2014bba}.
Measurements for $^2$H and $^3$He have been addressed, planned in some cases
but they have not been performed yet.

From the theoretical side,
$^4$He is a very important system: 
although really challenging, realistic calculations
are possible; besides, $^4$He is deeply bound and therefore
it represents the prototype of a typical finite nucleus;
in addition to that, it is
spinless, so that, experimentally, targets are easy to be implemented and 
data are easy to be analyzed.
Measurements were addressed and theoretical predictions 
proposed in Refs. \cite{Guzey:2003jh,Liuti:2005gi,Belitsky:2008bz}.
The first data for coherent DVCS off $^4$He have been 
recently published \cite{Hattawy:2017woc} 
and for the incoherent channel have been
already collected at JLab, by the EG6 experiment
of the CLAS collaboration, with the 6 GeV electron beam.
For the first time a
successful separation of coherent and incoherent contributions 
has been achieved.
A new impressive program is on the way at JLab12,
carried on by the ALERT collaboration 
\cite{Armstrong:2017wfw,Armstrong:2017zcm}.
In  Ref.
\cite{Hattawy:2017woc}, the
importance of new calculations has been addressed, for a
completely successful description of the collected data, not possible
with the models proposed long time ago, corresponding
in some cases to different kinematical regions.
New refined calculations are certainly important, above all,
for the next generation of accurate measurements.

Here, a conventional IA analysis of $^4$He GPD and
nuclear Compton form factor (CFF) is presented. 
The actual calculation is performed with basic nuclear
and nucleonic ingredients and the results are
compared with the recently published data 
\cite{Hattawy:2017woc}.

The paper is structured as follows.
In the second section, the formalism is introduced.
In the third one, nuclear and nucleonic ingredients
of the actual calculation are presented.
Then, numerical results are shown and discussed in the fourth section.
Eventually, conclusions and perspectives are given.

\section{Formalism}

\begin{figure}[h]
\hspace{-1.5cm}
\includegraphics[scale=0.40,angle=0]{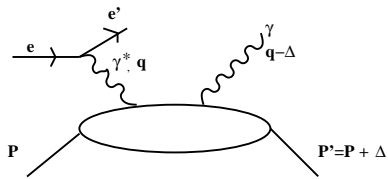}
\caption{
The generic coherent DVCS process off a target $A$.
}
\label{dvcsA}
\end{figure}

\begin{figure}[h]
\hspace{-1.5cm}
\includegraphics[scale=0.40,angle=0]{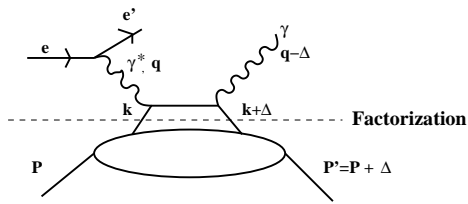}
\caption{
The handbag approximation to the process shown in Fig.
\ref{dvcsA}.
}
\label{dvcshb}
\end{figure}

The most general coherent DVCS process, 
$A(e,e'\gamma)A$, is shown in Fig. \ref{dvcsA}.
If the momentum transferred by the electrons, $Q^2$,
is much larger than $-t = -\Delta^2 = -(P-P')^2$, the momentum
transferred to the hadronic system
with initial (final) 4-momentum $P(P')$,
the hard vertex of the ``handbag'' 
diagram depicted in Fig. \ref{dvcshb} 
can be studied perturbatively, 
while the soft part, given by the blob
in the figure, is parametrized in terms of GPDs,
thanks to the factorization property 
demonstrated in Ref. \cite{Collins:1996fb}.

\begin{figure}[h]
\hspace{-1.5cm}
\includegraphics[scale=0.40,angle=0]{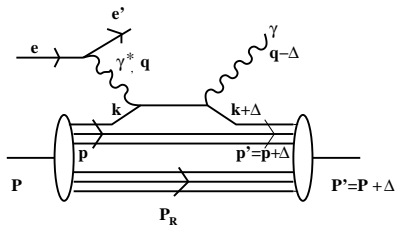}
\caption{
The Impulse Approximation description of the handbag process 
shown in Fig. \ref{dvcshb}.
}
\label{dvcsia}
\end{figure}

The formalism for DVCS off a
scalar target, exploiting only one
chiral even GPD 
at leading twist, has been developed in Ref. 
\cite{Belitsky:2008bz}.
In the following,  
a workable expression for 
$H_q^{^4He}$,
the  GPD of the quark of flavor
$q$ in the
$^4$He nucleus, will be derived within
the Impulse Approximation (IA) description of the handbag approximation,
depicted in Fig. \ref{dvcsia}.

From the expression of the leading twist light-cone correlator,
one can define $H_q^A$ for a generic scalar target
\cite{Diehl:2003ny}:

\begin{eqnarray}
H_q^A(x,\xi,\Delta^2)=(2P+\Delta)^+\int 
\frac{dr^-}{4\pi}e^{ix\bar{P}^+r^-}\langle P'|\bar{\psi}_q
\bigg(-\frac{r^-}{2}\bigg)\gamma^+ {\psi}_q
\bigg(\frac{r^-}{2}\bigg)|P \rangle 
\label{corr}\bigg|_{r^+=\vec{r}_\perp=0}~.
\end{eqnarray}

In the above equation, 
the dependence of
$H_q^A$ on three scalars is explicitly shown.
Besides $\Delta^2$, GPDs depend on
the skewness  $\xi =\frac{P^+-P'^+}{P^++P'^+}=-\frac{\Delta^+}{2\bar{P}^+} $,
i.e., the difference in plus momentum faction between the initial an 
the final states, completely
fixed by the external lepton kinematics, and  on $x$, 
the average plus momentum fraction of the struck parton 
with respect to the total momentum, not experimentally accessible.
 
The additional dependence of GPDs on the hard momentum scale $Q^2$
is not explicitly shown, for an easy presentation.
Here and in the rest of the paper,
the light-cone coordinates corresponding to
a generic 4-vector $v=(v_0, \vec v)$ are defined as 
$ v^\pm=\frac{v^0 \pm v^3}{\sqrt{2}} $. 

We remind that, in the case of zero momentum transfer, 
i.e. in the forward limit ($ P'=P $ , i.e $ \Delta^2=0, \xi=0 $), 
$H_q^{^4He}$ reduces to 
$^4$He parton distributions (PDF) accessed through DIS experiments

\begin{eqnarray}
H_q^{^4He}(x,0,0)= q^{^4He}(x)~,
\label{uno}
\end{eqnarray}

while its first moment 
yields the electromagnetic form factor of
$^4$He:

\begin{eqnarray}
\sum_q e_q \int_{0}^{1} dx \, H_q^{^4He}(x,\xi,\Delta^2)= 
F_C^{^4He}(\Delta^2)~,
\label{due}
\end{eqnarray}

where $e_q$ represents the charge of the quark of flavour $q$.

Besides, in the quark sector, one can define the plus 
momentum of the struck parton before and after the interaction:
\begin{eqnarray}
k^+ &=& (x+\xi)\bar{P}^+~, \\
k^{'+} &=&(k+\Delta)^+= (x-\xi)\bar{P}^+~, 
\end{eqnarray}
respectively. 
It is therefore clearly seen that $ x $ represents  
the average plus momentum fraction of the struck parton
with respect to the total nucleus momentum.

Now, the IA to the handbag approximation,
shown in Fig. \ref{dvcsia}, will be described.
The interacting parton, with momentum $k$, belonging to
a given nucleon with momentum $p$ in the nucleus, interacts with
the probe and it is afterwords reabsorbed, with 4-momentum
$k + \Delta$, by the same nucleon, without further re-scattering
with the recoiling three-body system.
One should notice that, in this scheme, only
nucleonic degrees of freedom occur explicitly
in the nuclear description.
In IA it is useful to rewrite the parton momenta 
also with 
respect to those of the inner nucleon $N$, as follows

\begin{eqnarray}
\label{par}
\xi'&=&-\frac{\Delta^+}{2p_N^+}~; \\
x'&=&\frac{\xi'}{\xi}x~.
\end{eqnarray}

The IA framework 
in the instant form of dynamics described in
Ref. \cite{Scopetta:2004kj}
for $^3$He is here extended to $^4$He.
The main steps are summarized here below.
Initially,
light-cone quantized states and operators are used.
The tensor product of
two complete sets of states can be inserted to the left and the 
right hand sides of the quark operator
in Eq. (\ref{corr});
the first set corresponds to the nucleon $ N $, supposed free, 
interacting with the virtual photon,
while the second set
describes the recoiling system, which consists of three fully
interacting particles.
Using the fact that the quark operator in Eq.
(\ref{corr})
is a one body operator, 
one can consider its action on the nucleonic degrees of freedom 
only. Separating the global motion from the intrinsic one, 
possible since at the end 
non-relativistic wave functions are used,
a convolution formula can be obtained
\begin{eqnarray}\label{gpd}
H_q^{^4He}(x,\xi,\Delta^2)
& = &
(2P+\Delta)^+\bigg[\int \frac{dr^-}{4\pi}e^{ix\bar{P}^+r^-}\bigg] dE \, 
\rho(E) \, 
\sum_{p_N \sigma \{\alpha\}} \langle P+\Delta|-p_N \, ,E \{\alpha\} ; 
p_N+\Delta \, , \sigma \rangle 
\nonumber
\\
& \times &
\langle p_N \,,\sigma\, ; p_N \, , E\, 
\{\alpha\}|P\rangle\bigg[\langle p_N+\Delta \, ,\sigma|\bar{q}
\bigg(-\frac{r^-}{2}\bigg)\gamma^+ q\bigg(\frac{r^-}{2}\bigg)|p_N ,
\sigma \rangle \bigg] 
\end{eqnarray} 
where the terms in the square brackets can be rearranged in terms of the 
generic light-cone correlator for the nucleon $ N $ considered for states with 
the same polarization $ \sigma $, that reads 
\cite{Diehl:2003ny}
\begin{equation}\label{corrn}
F_{++}^N=\sqrt{1-\xi^2}\bigg[ H_q^N-\frac{\xi^2}{1-\xi^2}E_q^N\bigg]~.
\end{equation}
In the above equation, 
in the kinematical region of the coherent channel of interest here,
the dominant term is given by the GPD $ H_q $.
Thus, in the following, we will consider only this contribution. 
Using Eq.\eqref{corrn} in Eq. \eqref{gpd} and properly
considering the partonic 
variables \eqref{par}, one arrives at a convolution formula

\begin{eqnarray}
H_q^{^4He}(x,\xi,\Delta^2)= \sum_N \int_{|x|}^1 { dz \over z } 
h_N^{^4He}(z,\Delta^2,\xi) \, H_q^{N} \left (
{x \over z} ,{\xi \over z}, \Delta^2
\right )
\label{conv}
\end{eqnarray}

between the GPD $H_q^N$ of the quark of flavor $q$ in the bound nucleon
$N$ and the off-diagonal 
light-cone momentum distribution of $N$ in $^4$He, which reads:

\begin{eqnarray}
h_N^{^4He}(z,\Delta^2,\xi) & = &
 \int dE \, \int d \vec p \,
P^{^4He}_N(\vec p, \vec p + \vec \Delta, E) \delta (z - \bar p^+/ \bar P^+)
\nonumber
\\
& = & 
 \frac{M_A}{M}\int dE \, \int_{p_{min}}^\infty  d p \,   \tilde{M} \, p \, P_N^{^4He}
(\vec{p},\vec{p}+\vec{\Delta},E)\delta\bigg(\tilde{z}\frac{\tilde{M}}{p}-
\frac{p^0}{p}-\cos\theta \bigg)~.
\label{hz}
\end{eqnarray}

In the last step of the above equation, we defined $ M(M_A) $ as the nucleon 
(nuclear) mass, $ \xi_A= \frac{M_A}{M}\xi $, $ \tilde{z}=z+\xi_A $ and  
$ \tilde{M} = \frac{M}{M_A} (M_A+\frac{\Delta^+}{\sqrt{2}}) $.  
The explicit form for the lower limit of integration in $ p=|\vec{p}| $ 
is given by 

\begin{eqnarray}
p_{min} (z,\Delta^2,\xi_A,E)=\frac{1}{2}\bigg|\frac{M_{A-1}^{*2}-
M_A^2(1-\frac{\tilde{M}}{M_A}\tilde{z})^2}{M_A(1-\frac{\tilde{M}}
{M_A}\tilde{z})}\bigg|~,
\end{eqnarray}

result obtained imposing the natural support for the function $\cos \theta$
in the argument of the delta function in Eq. (\ref{hz}),
with $ M_{A-1}^{*2} $ the squared mass of the final $ A-1 $-body excited 
states. 

The off-diagonal light cone momentum
distribution of the nucleon $N$ in $^4$He
is defined through its non-diagonal spectral function:

\begin{eqnarray}
P_N^{^4He}(\vec p, \vec p + \vec \Delta,E) 
& = &
\rho(E) \sum_{\{\alpha\} \sigma}\langle P+\Delta|-p_N E\alpha ,p_N+ 
\Delta, \sigma \rangle \langle p_N \sigma, - p_N E \alpha|P\rangle
\\
& = &
n_0( \vec p, \vec p
+ \vec \Delta)  \delta(E)
+ P_1(\vec p, \vec p + \vec \Delta, E)~,
\label{ndspec}
\end{eqnarray}
being $ \rho(E) $ the energy density for the final states.
The overlaps appearing in this formula include wave functions of the recoiling
three-body system, which can be a bound system, 
a two-body or a three-body
scattering state with any possible relative energy between the constituents.  
We reiterate that
any interaction of the debris originating by the struck nucleon with the 
remnant $(A-1)$ nuclear system is instead disregarded, as usual
in the IA scheme.

The forward limit of the expression Eq.
(\ref{ndspec})
leads to the one-body diagonal spectral function of $^4$He,
$P_N^{^4He}(\vec p,E)$, so that Eq. (\ref{hz})
reduces to

\begin{eqnarray}
h_N^{^4He}(z,0,0)= f_N^{^4He}(z) = \int dE \, \int d \vec p \,
P_N^{^4He}(\vec p, E) \delta \bigg(z - \frac{\sqrt{2}p^+}{M}\bigg)~.
\label{fz}
\end{eqnarray}

Using this result, Eq. (\ref{conv}) reproduces 
in the forward limit the correct
IA result for the nuclear PDF 
(see, e.g., Ref. \cite{CiofidegliAtti:1990dh}),
in agreement with Eq. (\ref{uno}).

Besides, the $x-$ integral of Eq. (\ref{conv}) yields formally 
the IA, one-body approximation to the nuclear form factor,
so that the constraint Eq. (\ref{due}) is also formally
fulfilled. 

A few caveats have to be addressed:

$i)$ in the present Instant Form calculation the number
of particle sum rule and the momentum sum rule cannot
be fulfilled at the same time.
In particular, the momentum sum rule is here
violated by a few percent. To overcome this drawback a Poincar\`e
covariant Light-Front approach
could be used. Relevant steps
towards this goal have been done 
for a three-body nuclear target \cite{DelDotto:2016vkh}.

$ii)$ the present scheme is not covariant and, as a consequence,
the GPDs, although scalar, turn out to be frame dependent.
For GPDs calculation, as well as for form factors, 
at high momentum transfer the use of LF dynamics would be 
the proper framework.
Nevertheless, in the experiment
discussed in the present paper, the momentum transfer is rather low
and we found that, in the observables we are going to show,
the results in the Laboratory frame or in the Breit frame 
differ at most by a few parts in one thousand.
Therefore, at the moment, this problem is not a numerically relevant one. 
The results presented later on have been obtained in the Laboratory frame.


\begin{figure}[t]
\hspace{-1.5cm}
\includegraphics[scale=0.40,angle=270]{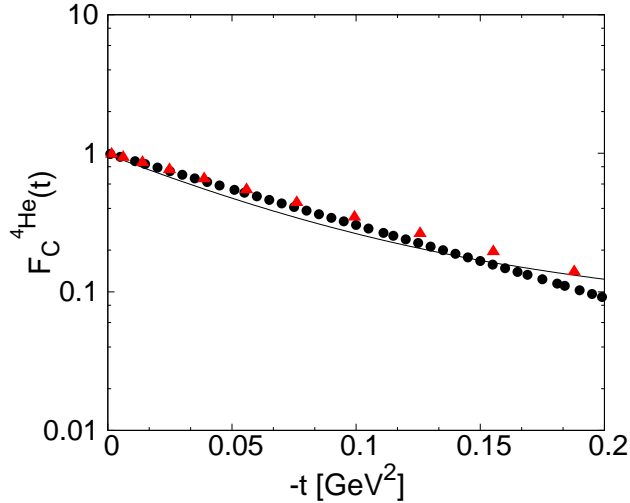}
\caption{
(Color online)
The $^4$He form factor obtained as the integral
of the $^4$He GPDs calculated in the present approach, 
Eq. (\ref{conv}) (full), 
compared with data at low $t$ (dots) 
\cite{Ottermann:1985km}, the ones
relevant for the discussion presented here.
Red triangles represent the one-body part
of the Av18+UIX calculation of the form factor
shown in Ref. \cite{Camsonne:2013dfp} (see text).
}
\label{ff}
\end{figure}

\begin{figure}[t]
\hspace{-1.5cm}
\includegraphics[scale=0.40,angle=270]{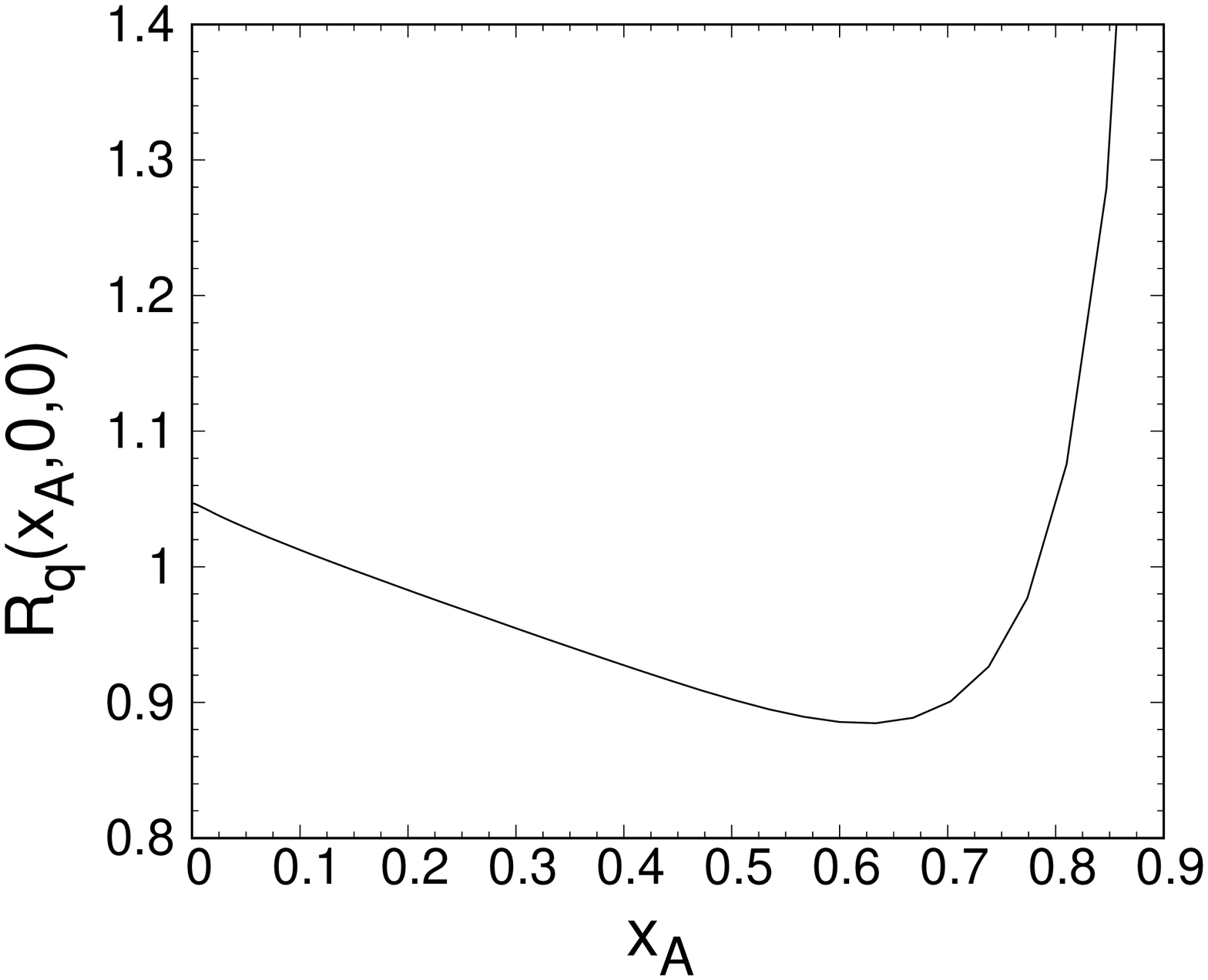}
\caption{The ratio Eq. (\ref{rat})}
\label{emc}
\end{figure}


\begin{figure}[h]
\hspace{-1.5cm}
\includegraphics[scale=0.40,angle=270]{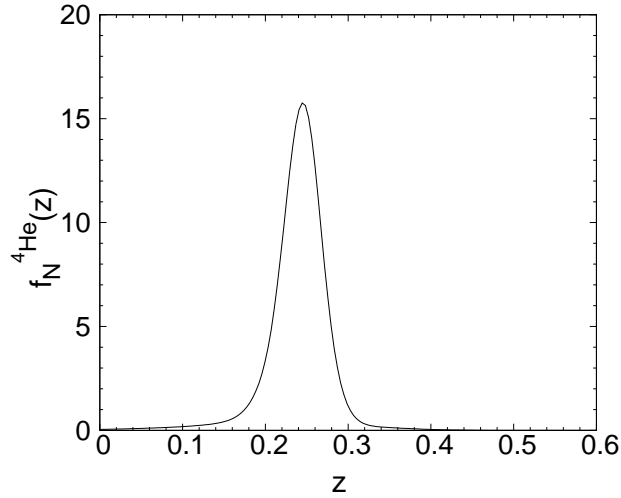}
\caption{The light-cone momentum distribution
for the nucleon $N$ in $^4$He, 
Eq. (\ref{fz}).}
\label{lc}
\end{figure}

\begin{figure}[t]
\hspace{-1.cm}
\includegraphics[scale=0.28,angle=270]{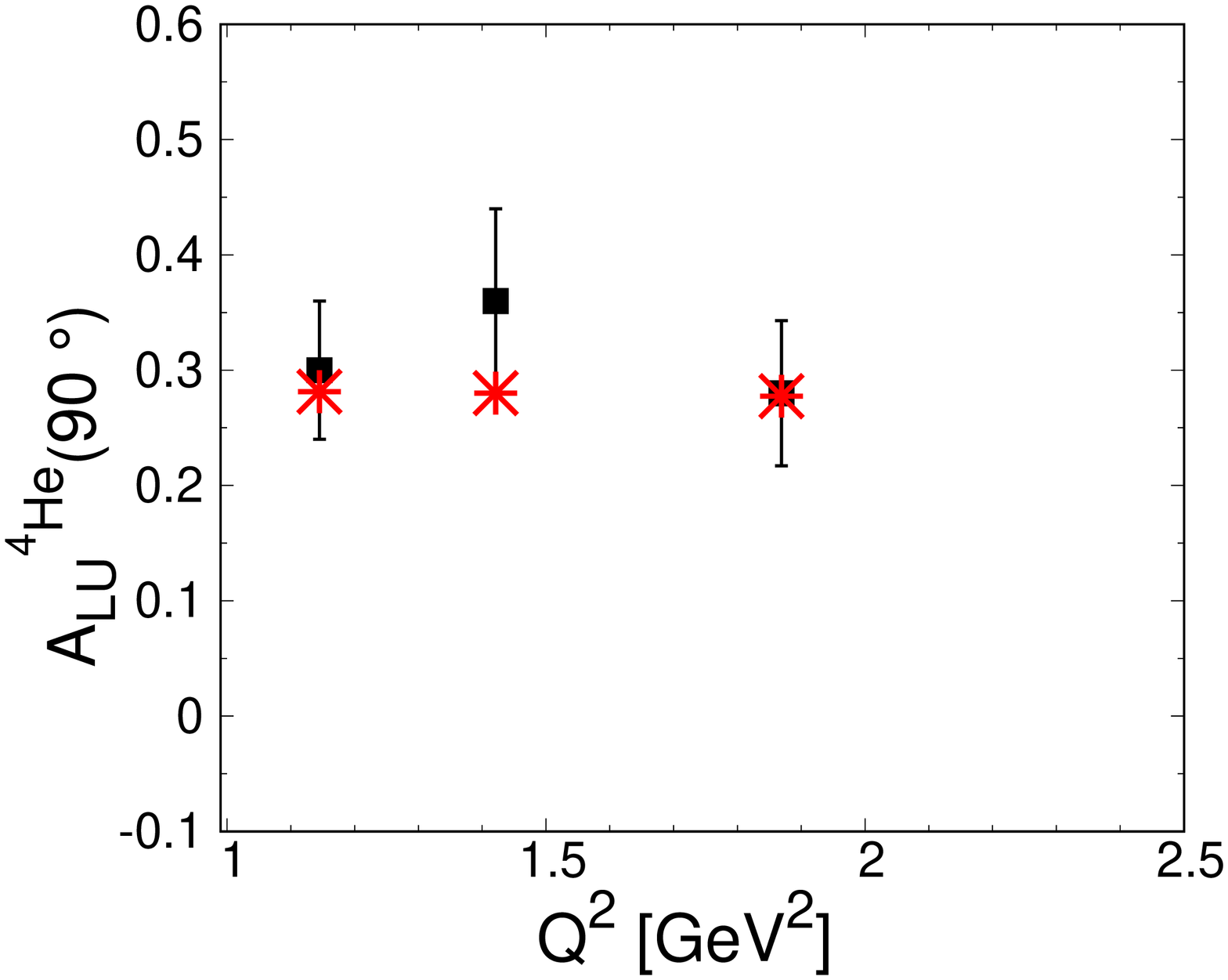}
\hspace{-1.5cm}
\includegraphics[scale=0.28,angle=270]{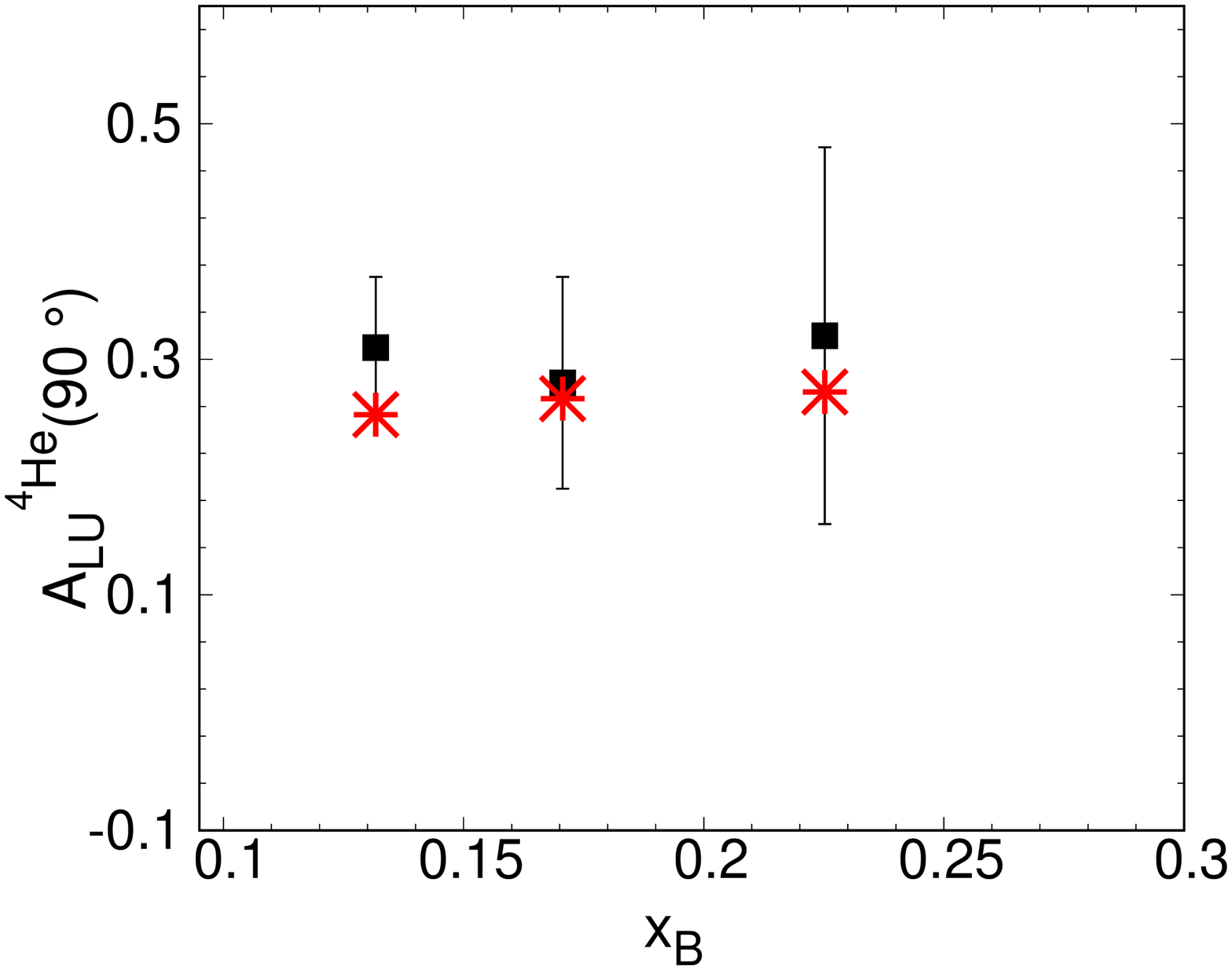}
\hspace{-1.5cm}
\includegraphics[scale=0.28,angle=270]{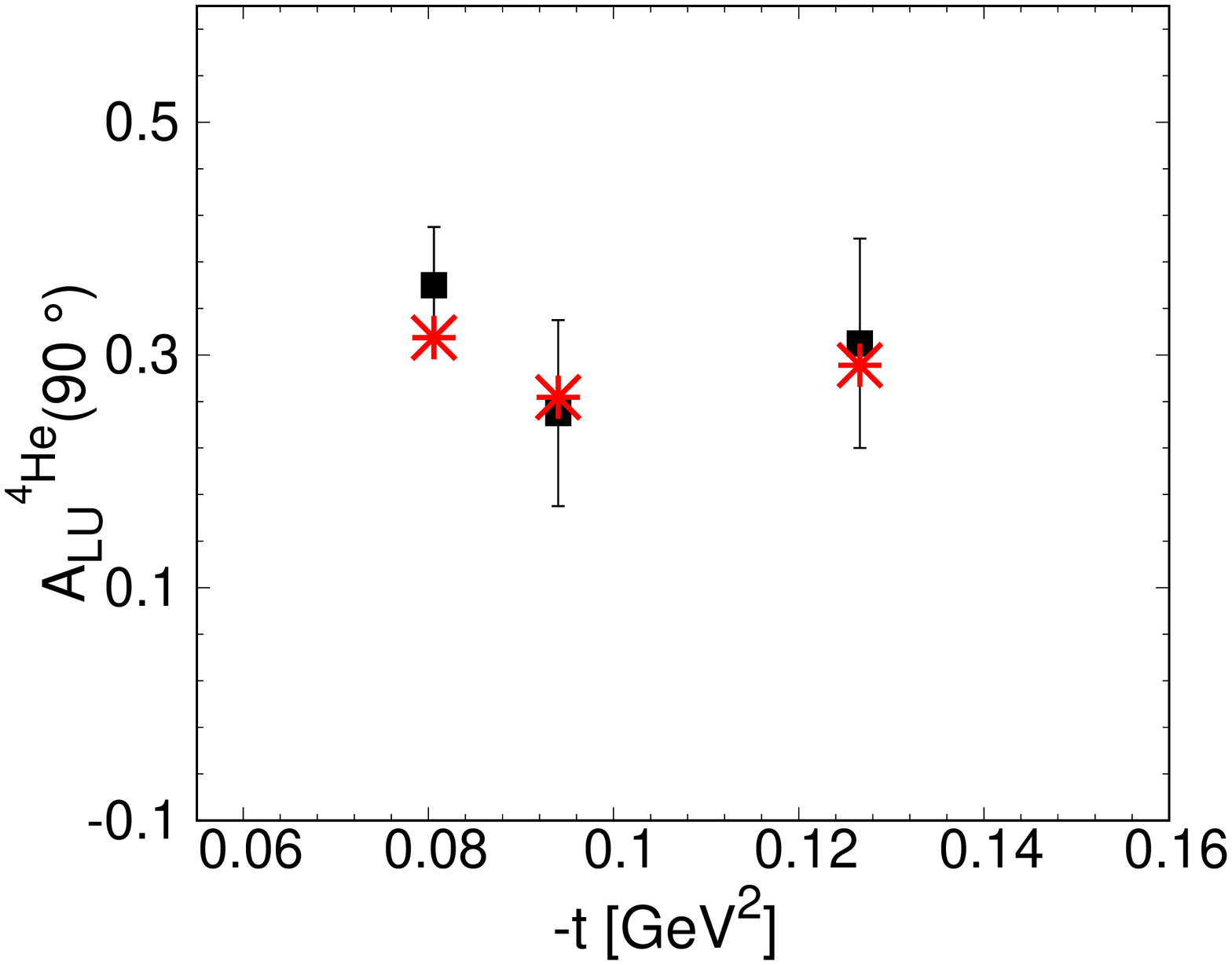}
\caption{(Color online) $^4$He azimuthal beam-spin asymmetry $A_{LU}(\phi)$,
for $\phi = 90^o$: results of this approach (red stars) compared with data
(black squares)
\cite{Hattawy:2017woc}.
From left to right, the quantity is shown in the experimental
$Q^2$, $x_B$ and $t$ bins, respectively. 
}
\label{alu}
\end{figure}

\begin{figure}[h]
\hspace{-1.cm}
\includegraphics[scale=0.28,angle=270]{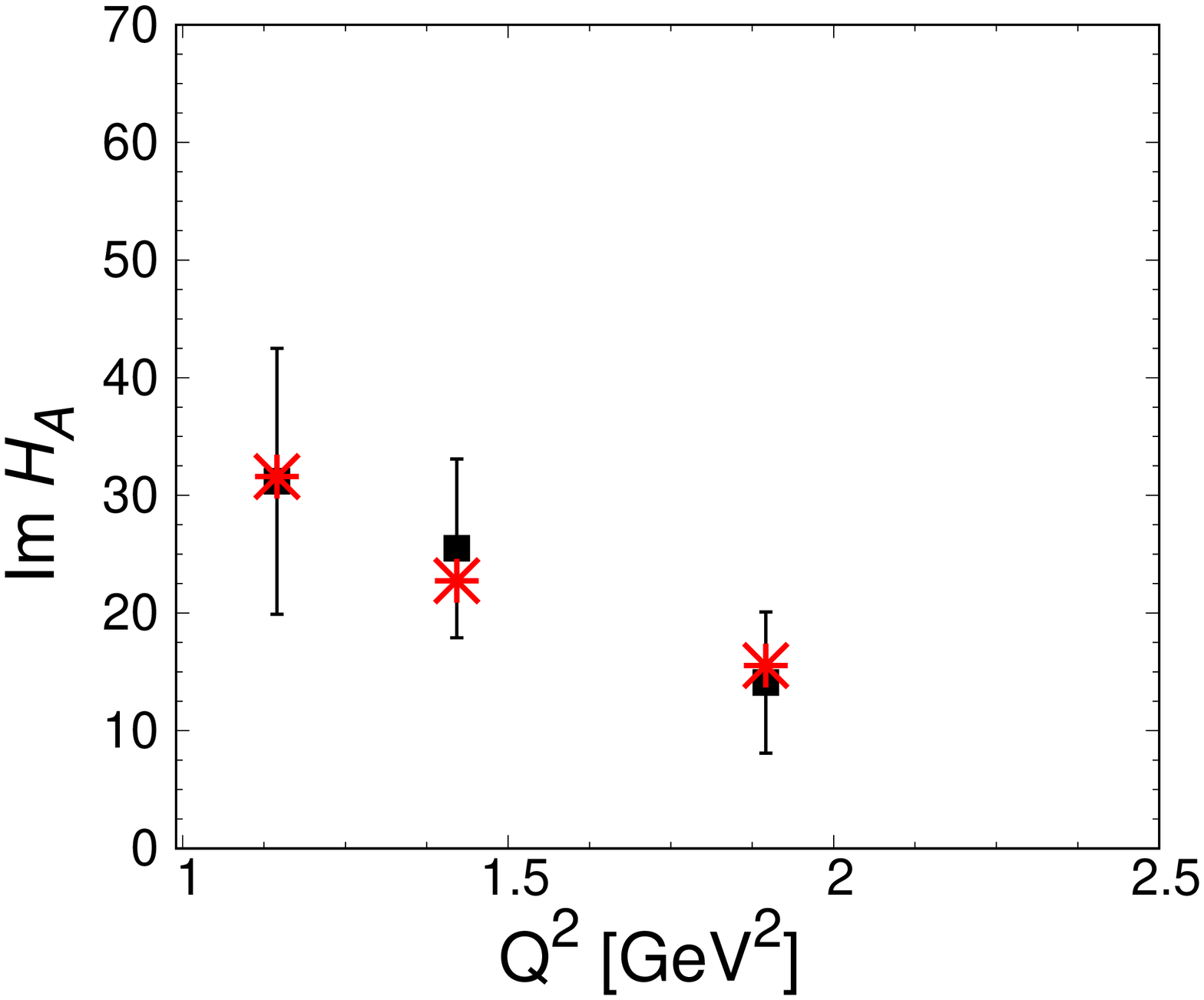}
\hspace{-1.5cm}
\includegraphics[scale=0.28,angle=270]{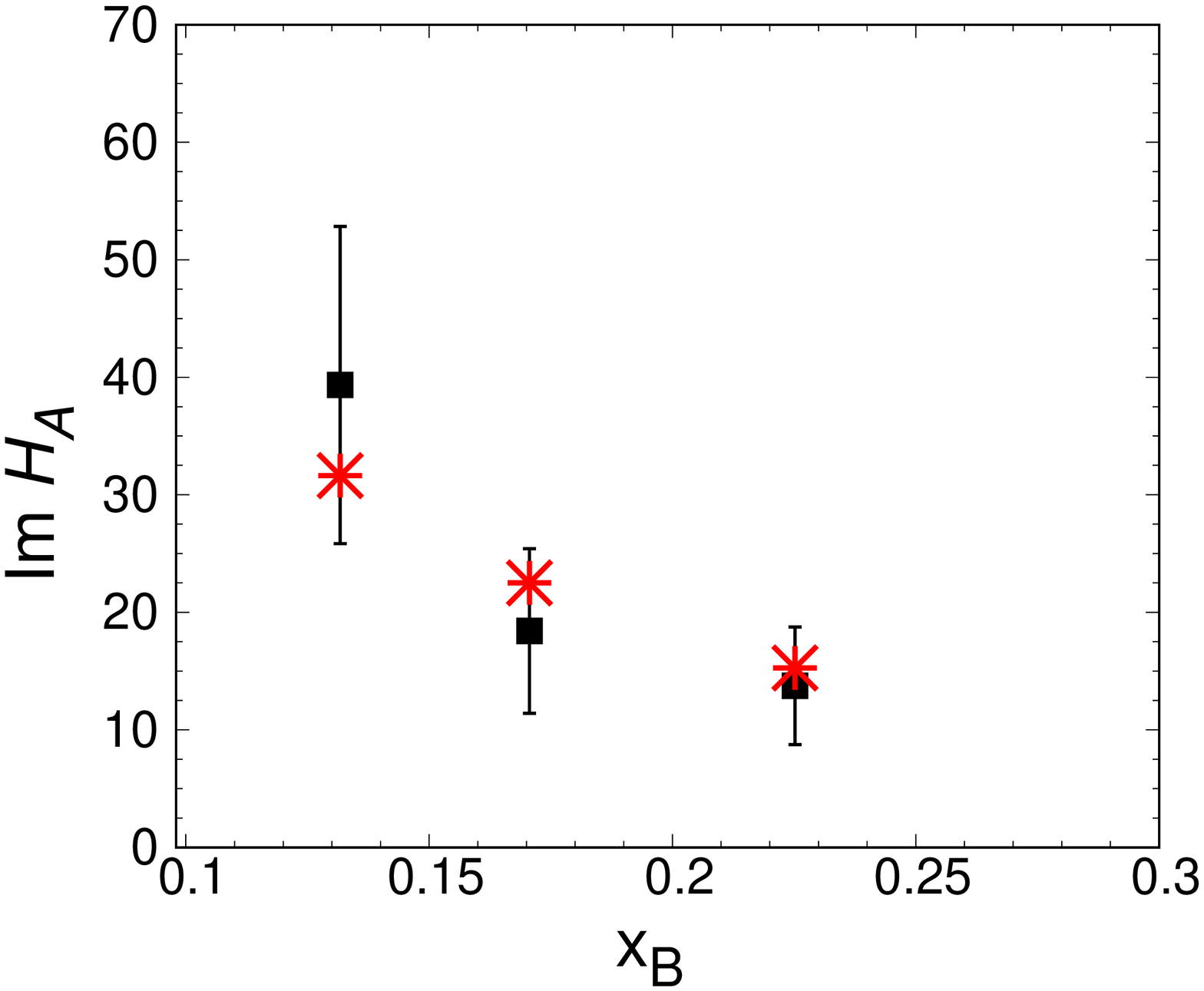}
\hspace{-1.5cm}
\includegraphics[scale=0.28,angle=270]{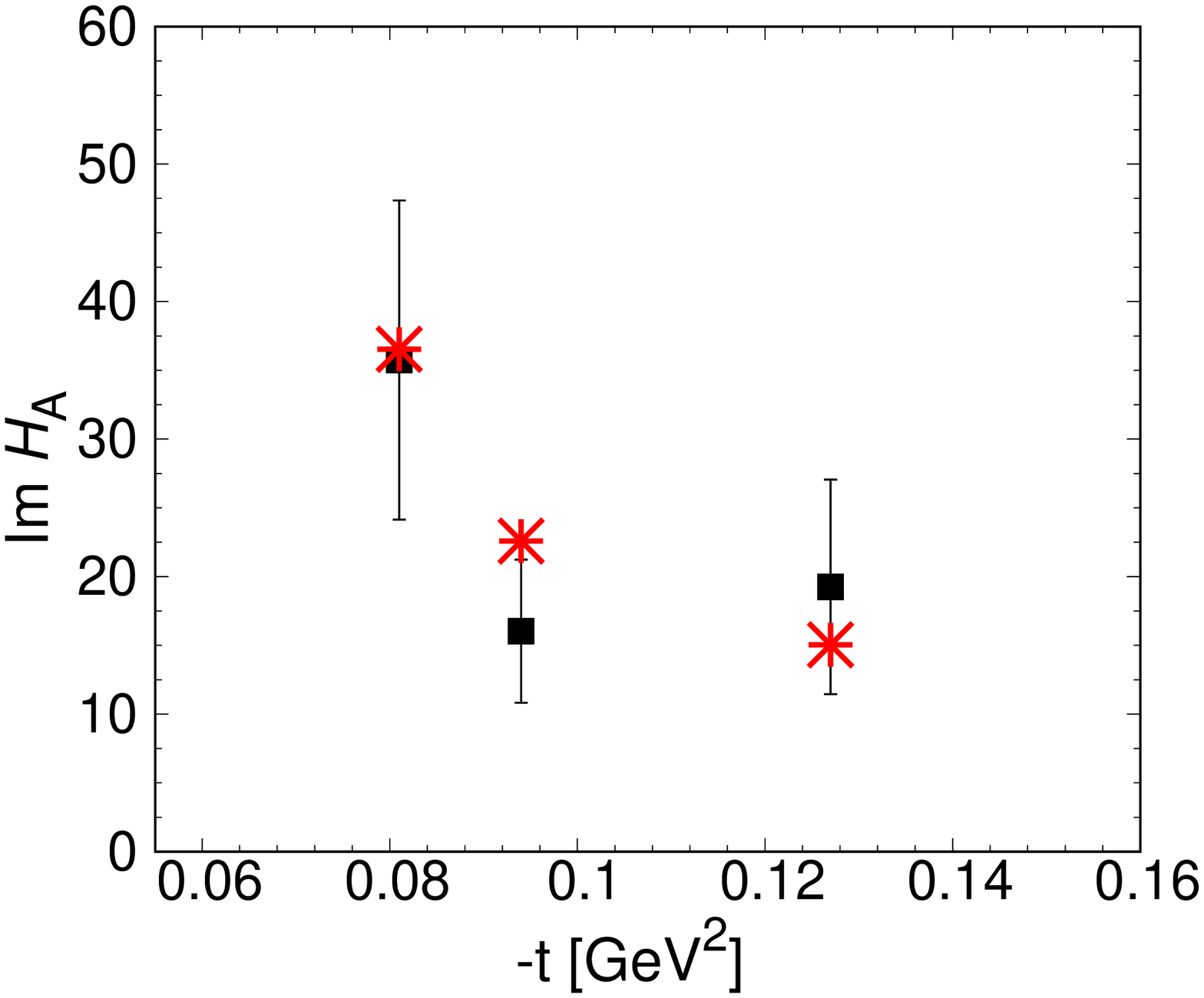}
\caption{(Color online) The imaginary part of the Compton form factor
for $^4$He: results of this approach (red stars) compared with data
(black squares)
\cite{Hattawy:2017woc}.
From left to right, the quantity is shown in the experimental
$Q^2$, $x_B$ and $t$ bins, respectively. 
}
\label{imh}
\end{figure}

\begin{figure}[h]
\hspace{-1.cm}
\includegraphics[scale=0.28,angle=270]{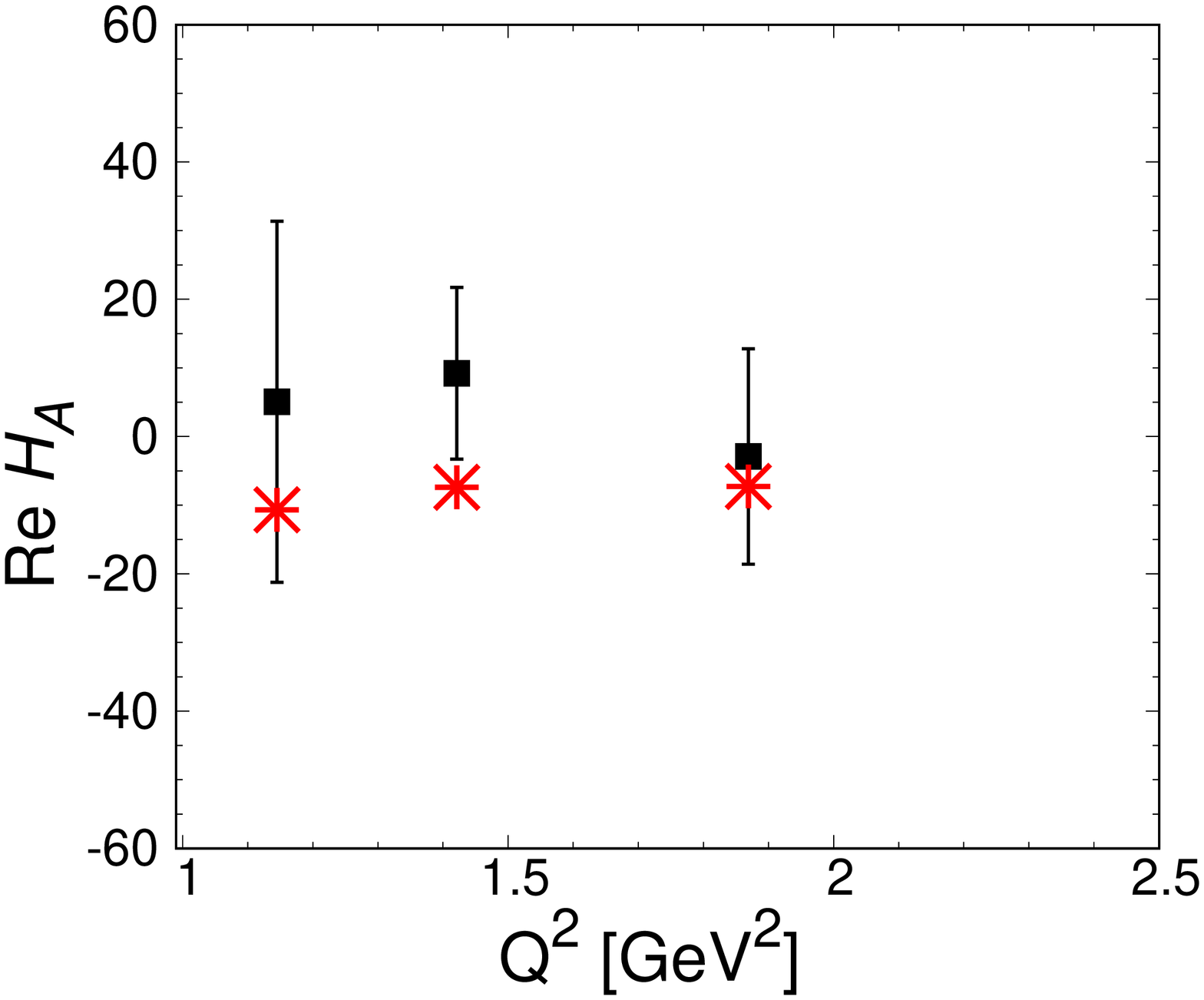}
\hspace{-1.5cm}
\includegraphics[scale=0.28,angle=270]{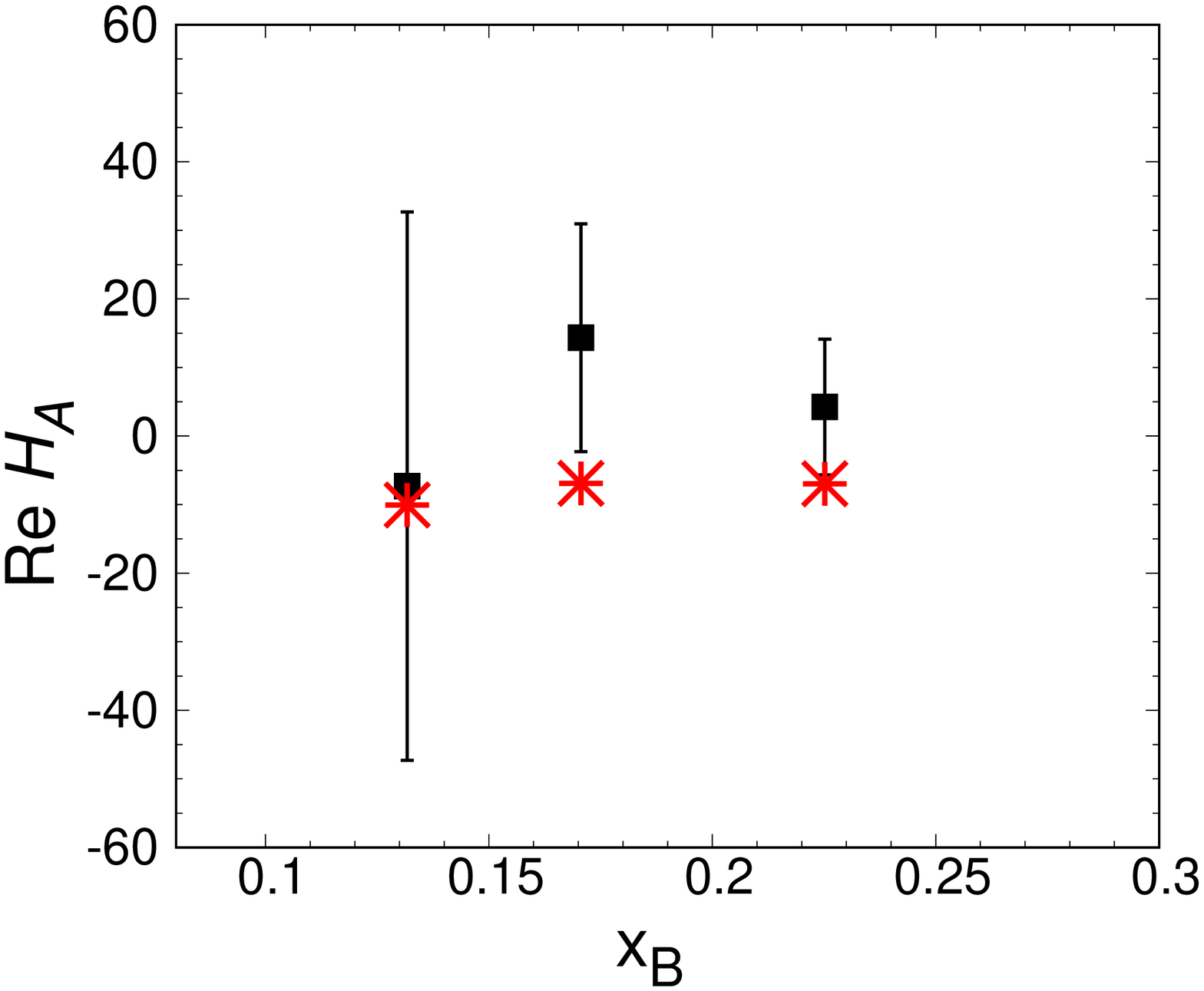}
\hspace{-1.5cm}
\includegraphics[scale=0.28,angle=270]{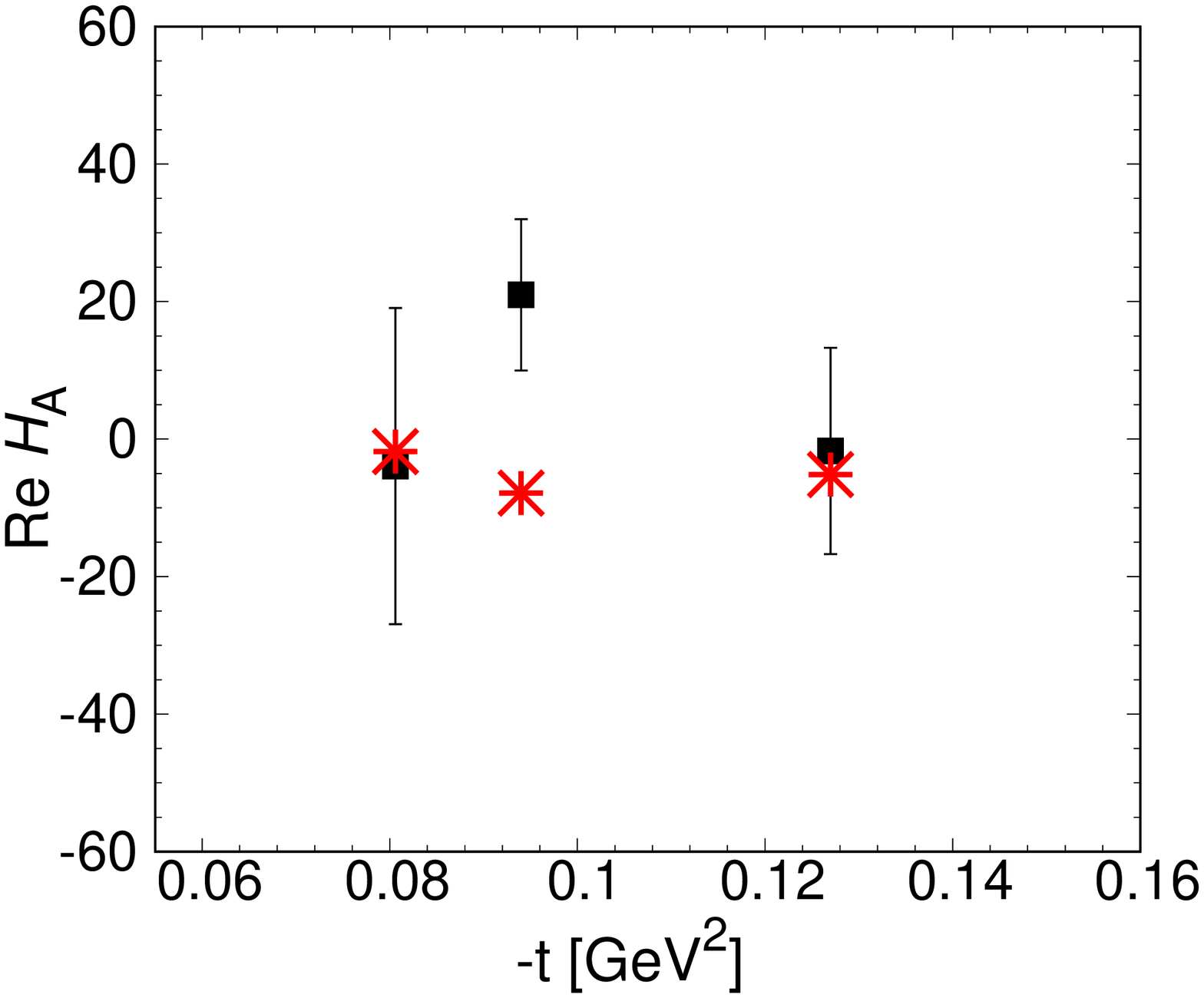}
\caption{(Color online) The real part of the Compton form factor
for $^4$He: results of this approach (red stars) compared with data
(black squares)
\cite{Hattawy:2017woc}.
From left to right, the quantity is shown in the experimental
$Q^2$, $x_B$ and $t$ bins, respectively. 
}
\label{reh}
\end{figure}

Concluding this section, one should notice that, 
in the present IA approach, the
momentum scale $ Q^2$ of the nuclear GPD 
is entirely given by that of the nucleon GPD and, 
for the sake of a readable presentation, 
it is not explicitly written in the following.

\section{Set up of the calculation}

It is clear from the previous section that, in order to actually
evaluate
the $^4$He GPD and then the cross section for
coherent DVCS off $^4$He, we need an input
for the nuclear non-diagonal spectral function
and for the nucleonic GPD.

Concerning the nuclear part,
only old attempts exist of obtaining a spectral
function of $^4$He
\cite{Morita:1991ka}.
A realistic description of the
two and, above all, three-body scattering states 
in the recoiling system is a really complicated
few-body problem. Moreover, one would need here
a non-diagonal spectral function, a quantiy rather more complicated
than the diagonal one.

We have planned a full realistic calculation
of the $^4$He spectral function; 
in this work, use of the following model has been done:

\begin{eqnarray}
P^{^4He}_N(\vec p, \vec p + \vec \Delta, E) 
& = &
n_0( \vec p, \vec p
+ \vec \Delta)  \delta(E^*)
+ P_1(\vec p, \vec p + \vec \Delta, E^*)
\nonumber
\\
& = & 
n_0( |\vec p|, |\vec p
+ \vec \Delta|, \cos \theta_{\vec p, \vec p + \vec \Delta } )  \delta(E^*)
+ P_1(|\vec p|, |\vec p + \vec \Delta|, 
\cos \theta_{\vec p, \vec p + \vec \Delta }, E^*)
\nonumber
\\
& \simeq & {a_0(|\vec p|) a_0(|\vec p
+ \vec \Delta|) } \delta(E^*)
+ n_1(| \vec p|,|\vec p + \vec \Delta|)  \delta(E^* - \bar E)
\label{pkemodel}
\end{eqnarray}

with the removal energy $E=|E_A| - |E_{A-1}|+E^*$ defined in terms
of the ground state binding energies of $^4$He and of 
the recoiling three nucleon 
system, $E_A$ and $E_{A-1}$, respectively, and in terms 
of the excitation energy
of the recoiling system, $E^*$. Besides, one has
$n_1(| \vec p| ) = n( | \vec p| ) - n_0( | \vec p| )$, and

\begin{equation}
n_0( |\vec p|) = | a(|\vec p|) |^2
\end{equation}

with $a(|\vec p|)$ the overlap of the wave functions
of the 4- and 3-body bound systems:

\begin{equation}
a(|\vec p|) = < \Phi_3(1,2,3)\chi_4 \eta_4 | j_0( | \vec p | R_{123,4})
\Phi_4(1,2,3,4) >~.
\end{equation}

In our calculation, 
$n_0(k)$, the momentum distribution
with the recoiling system in the ground state, and 
the total momentum distribution $n(k)$ have been evaluated 
using variational wave functions
for the 4-body \cite{PisaWF} and 3-body \cite{PisaWF3} systems
obtained through the hyperspherical 
harmonics method \cite{hh},
within the Av18 NN interaction \cite{Wiringa:1994wb},
including UIX three-body forces \cite{Pudliner:1995wk}.

The spirit of the approximation Eq. (\ref{pkemodel}) is the following.
In the first line of the equation, the rotational invariance of the problem
has been exploited, showing a dependence on the absolute values
of the initial and final momentum of the struck nucleon, and on
the angle between these two momenta. In the second line, the so-called closure
approximation to the spectral function is used in the excited sector
described by the spectral function $P_1$, i.e., an average value of the 
removal energy is chosen so that the non-diagonal spectral function
reduces to a non-diagonal momentum distribution. 
The average value $\bar E$ of the excitation energy $E^*$
of the recoiling system
is evaluated through the model diagonal spectral function, 
based on the same Av18+UIX interaction, proposed
in Ref. \cite{Viviani:2001wu,Rinat:2004ia},
representing a realistic 
update of the one presented in \cite{CiofidegliAtti:1995qe}.
In the last step, also the angular dependence is disregarded, so that
the non-diagonal momentum distributions can be modeled
on the basis of the known diagonal ones.

For the nucleonic part,
the well known GPD model elaborated by Goloskokov and Kroll (GK) 
\cite{Goloskokov:2007nt,Goloskokov:2011rd} has been used.
We remind here, for the reader convenience, its main features.
The explicit form of GPDs is obtained fitting high 
energy Deeply Virtual Meson production (DVMP) data. 
This guarantees the access to the low $ x $ region. 
The structure of the $ (x,\xi) $ dependence
is built through the Double
Distributions representation 
\cite{Radyushkin:1998es}, so that 
the polinomiality property is automatically satisfied, while
the $ t $-dependence is parametrized  
using a Regge-inspired profile function.
The model is valid in principle at $Q^2$ values larger than those of 
interest here, in particular at $Q^2 \ge 4$ GeV$^2$.

\section{Numerical results}

With the ingredients presented in the previous section at hand, 
a numerical evaluation of the nuclear GPD Eq. \eqref{conv} 
is possible and a comparison with the related
experimental observables, recently accessed by the EG6
experiment at JLab, can be performed. 
Before than that, let us consider two useful numerical tests
of the formalism.

First of all, one should recover the IA result for the
electromagnetic form factor (ff)
(for example, the one-body result in Ref.
\cite{Camsonne:2013dfp}),
by $x-$integration
of the obtained GPD using Eq. (\ref{uno}):
\begin{align}\label{506}
\frac{1}{2}\sum_q e_q \int dx \, H_q^{^4He}(x,\xi,\Delta^2)&= \frac{1}{2}
\sum_{N,q} e_q \int_0^1 dx \int_x^1 \frac{d z}{z}
h_N^{^4He}(z,\xi,\Delta^2)
H_q^N \left (\frac{x}{z},\frac{\xi}{z},\Delta^2
\right )
\nonumber
\\ &=\frac{1}{2}\sum_{N,q}
e_q
\int_0^1 d \left ( {x \over z} \right ) \, H_q^N \left (
{x \over z}, {\xi \over z} ,\Delta^2
\right ) \int_0^1 d z h_N^{^4He}(z,\xi,\Delta^2) 
\nonumber
\\ &=\frac{1}{2}\sum_{N,q} 
F_q^N(\Delta^2)F_N^{^4He}(\Delta^2)=F^{^4He}(\Delta^2)~.
\end{align}

Let us notice that the factor of 2 in the denominator
of the above equation, 
i.e. the charge of the nucleus under scrutiny in units of $e$, 
guarantees the standard normalization $ F^{^4He}(0)=1 $. 
This quantity is shown in Fig. \ref{ff}.
Despite of the approximated $\Delta$ dependence of the
spectral function described in the previous section,
reasonable agreement with the data 
\cite{Ottermann:1985km} is obtained
for the low values of $(-t)$
accessed by the EG6 experiment at JLab.
The agreement 
has certainly to be improved, evaluating
a realistic spectral function of $^4$He, 
for a precise description
of the accurate data of the next generation of measurements.
The size of the target is reproduced with good accuracy.
Quantitatively, we get $\sqrt{<r^2_{rms}>}\simeq$ 1.80 fm,
to be compared with the experimental value 1.671(14) fm \cite{Ottermann:1985km}.
In Fig. \ref{ff}, for completeness, also the results for the nuclear
form factor obtained within a one-body Av18+UIX calculation,
compared with data in Ref. \cite{Camsonne:2013dfp}, have been shown.
Within a realistic Av18+UIX spectral function, one would have
obtained this kind of results for the nuclear ff.
We stress anyway that the direct calculation of the $^4$He ff
requires only the wave function of the bound state, while
the calculation through GPDs, performed here
as a check, requires all the wave functions of the 
spectral decomposition of $^4$He.

As a second test, we checked that
the obtained GPD has the expected forward limit.
This is seen in Fig. \ref{emc}, where the ratio

\begin{eqnarray}
R_q^{^4He}(x_A)= H_q^{^4He}(x_A,0,0)/H_q^{N}(x_A,0,0)~
\label{rat}
\end{eqnarray}

is shown as a function of $x_A= M_A/M x \simeq 4 x$, to have
an easy comparison with the results shown in the literature of
DIS phenomena.
In the above equation,
the numerator is given by the forward limit of Eq. (\ref{conv})
and the denominator by the forward limit of the 
model used for the nucleon GPD. No relevant difference is
found between the result for $q=u$ and $q=d$, as it is natural for
an isoscalar nucleus.
The typical EMC-like behavior found for this ratio in IA is reproduced.
One should notice anyway that the true EMC ratio
is defined dividing the nucleus $F_2$ structure functions
by the same quantity for the deuteron, while the quantity shown
here is obtained in terms of parton distributions of a given flavor.
This behavior is therefore related to the EMC effect but it represents
a different quantity.

The results of checks 1) and 2) are therefore rather encouraging.

Size and relevance of nuclear effects can be inferred from the
behaviour of the light-cone momentum distribution Eqs. 
(\ref{hz}) and (\ref{fz}).
If nuclear effects were negligible, these functions would be 
delta functions.
The light-cone momentum distribution, in the forward limit,
is shown in Fig. \ref{lc}.
One can see in passing that the present approach predicts a vanishing
DVCS cross section already for $\xi$ as small as 0.15, representing
the width of the shown distribution.
Indeed, $\xi$ is the fraction of plus momentum transfer and cannot
exceed the width of $f(z)$, if we want the target to be 
intact after the interaction. 
If, in future measurements, coherent DVCS were observed
at larger value of $\xi$, the
role of non-nucleonic degrees of freedom would be
exposed, as suggested in the seminal paper \cite{Berger:2001zb}.

Now, the comparison of our results
with the data of the EG6 experiment is eventually
performed.

In the EG6 experiment the crucial measured
observable is the single-spin asymmetry
$A_{LU}$, which can be 
extracted from the reaction yields for the two electron
helicities ($N^{\pm}$):
\begin{equation}
A_{LU} = \frac{1}{P_{B}} \frac{N^{+} - N^{-}}{N^{+} + N^{-} },
\end{equation}
where $P_{B}$ is the degree of longitudinal polarization of the incident 
electron beam.
The DVCS amplitude depends on the GPDs. In EG6 kinematics, 
the cross section of real photon 
electroproduction is dominated by the BH contribution, while the DVCS 
contribution is very small. However, the DVCS contribution is
enhanced in the observables sensitive to the interference term, {\it e.g.} 
$A_{LU}$. The three terms entering the cross section calculation,
the squares of the BH and DVCS amplitudes and their interference term, 
depend on the
azimuthal angle $\phi$ between the $(e,e^\prime)$ and 
$(\gamma^*,^4$He$^\prime)$ planes,
as shown for the nucleon in Ref.~\cite{Belitsky:2001ns} and for 
the spin-zero targets
in Refs.~\cite{Kirchner:2003wt,Belitsky:2008bz}. Based on this work, $A_{LU}$ 
for a spin-zero hadron can be expressed at leading-twist as
\begin{equation}
A_{LU}(\phi) = 
\frac{\alpha_{0}(\phi) \, \Im m(\mathcal{H}_{A})}
{\alpha_{1}(\phi) + \alpha_{2}(\phi) \, \Re e(\mathcal{H}_{A}) 
+ \alpha_{3}(\phi) \, 
\big( \Re e(\mathcal{H}_{A})^{2} + \Im m(\mathcal{H}_{A})^{2} \big)}.
\label{eq:A_LU-coh}
\end{equation}
Explicit forms for the kinematic factors $\alpha_i$ are derived 
from expressions in
Ref.~\cite{Belitsky:2008bz} and are functions of Fourier harmonics in the 
azimuthal angle $\phi$, the nuclear form factor $F_A(t)$ and kinematical
factors. Using the different $\sin(\phi)$ and $\cos(\phi)$ contributions, 
in the experimental analysis,
both the imaginary and real parts 
of the so-called Compton Form Factor $\mathcal{H}_{A}$ have been extracted 
by fitting the $A_{LU}(\phi)$ distribution.
In turn, the imaginary and real parts 
of $\mathcal{H}_{A}$ are defined as follows
\cite{Guidal:2013rya}:
\begin{eqnarray}
\Im m(\mathcal{H}_{A}) & = & H_A(\xi,\xi,t)-H_A(-\xi,\xi,t),
\label{eqim}
\\
\Re e(\mathcal{H}_{A}) & = & \mathcal{P} 
\int_{0}^{1}dx[H_A(x,\xi,t)-H_A(-x,\xi,t)] \, C^{+}(x,\xi), 
\label{eqre}
\end{eqnarray}
in terms of the nuclear GPD $H_A$, where $\mathcal{P}$ is 
the Cauchy principal value integral, and a coefficient function $C^{+}= 
\frac{1}{x-\xi} + \frac{1}{x+\xi}$ has been introduced.

Using our result for the GPD of $^4$He, Eq. (\ref{conv}), we could
evaluate Eqs. (\ref{eq:A_LU-coh}), (\ref{eqim}) and (\ref{eqre}).
Results are reported in Figs. \ref{alu}, \ref{imh} and \ref{reh},
respectively, compared with the EG6 data.

In Fig.~\ref{alu}, $A_{LU}$ is shown at  
$\phi=90^o$ as a function
of the kinematical variables $Q^2$, $x_B=Q^2/(2 M \nu)$, 
and $t$. Due to limited 
statistics, in the experimental analysis
these latter variables have been studied separately 
with a two-dimensional 
data binning. The same procedure has been used in our theoretical estimate.
For example, each point at a given $x_B$ has been obtained
using for $t$ and $Q^2$ the corresponding average experimental values. 
Overall, a very good agreement is found.

The same happens for $\Im m(\mathcal{H}_{A})$
shown in Fig.~\ref{imh}, while for $\Re e(\mathcal{H}_{A})$
the agreement is somehow less satisfactory, as it is seen
in Fig.~\ref{reh}.
In particular, one point in the $t$ dependence
is not reproduced. One should not
forget anyway that the present data do not constraint enough
$\Re e(\mathcal{H}_{A})$, a quantity appearing multiplied by small coefficients
in Eq. (\ref{eq:A_LU-coh}).

The Cauchy principal value integral in Eq. (\ref{eqre}) has been
evaluated numerically using both the standard Cern library routines 
and the procedure described in Ref. \cite{ausinc},
obtaining a negligible difference with the two methods.
From the theoretical side we note also that the result 
for $\Re e(\mathcal{H}_{A})$
is strongly
dependent on the model used to evaluate
the nucleon GPD in the convolution formula.
We also note that the GK model is supposed to work properly at $Q^2 > 4$
Ge$V^2$. Here we have forced its validity at much lower $Q^2$ values,
with remarkable success.

On the light of this comparison, we can conclude that
the description of the present data does not require
exotic arguments, such as dynamical off-shellness.
As a matter of fact, our calculation shows that a careful
use of basic conventional ingredients is able to
reproduce the data.

\section{Conclusions and perspectives}

A thorough analysis of the available data on coherent
deeply virtual Compton scattering off $^4$He has been presented.
The framework is the impulse approximation description
of the process at leading twist, given by the handbag contribution.
In this way, a convolution formula is obtained,
in terms of a non-diagonal one-body spectral function 
of the nucleus and the GPD of the bound nucleon.
The nucleonic contribution is parametrized through the Goloskokov-Kroll 
model. The nuclear part is given by a model of the one-body
non diagonal spectral function, which reproduces in the proper limit
the exact Av18+UIX diagonal momentum distribution. A reasonable
description of the electromagnetic form factor at the low values
of the momentum transfer, relevant for the specific experimental kinematics,
is reproduced. In the forward limit, the nuclear parton distributions
show the expected emc-like behaviour. 
Overall very good agreement is found for the observables
recently measured at Jefferson Lab.
As a matter of facts, our calculation shows that a careful
analysis of the reaction mechanism in terms of
basic conventional ingredients is able to
describe the data. 
We can conclude that the present experimental accuracy does not require
the use of exotic arguments, such as dynamical off shellness.
Nevertheless, a serious benchmark calculation in the kinematics of the
next generation of precise measurements at high luminosity will require
an improved treatment of both the nucleonic and the nuclear
parts of the calculation. The latter task includes the realistic evaluation
of a one-body non diagonal spectral function of $^4$He.
Work is in progress towards this challenging direction.   
In the meantime, the straightforward approach proposed here
can be used as a workable framework for the planning of future
measurements.

\section*{Acknowledgments}
We warmly thank R. Dupr\'e and M. Hattawy for many
helpful explanations on the EG6 experiment, and L.E. Marcucci
for sending us the results for the one body 
form factor calculation within the AV18 plus UIX potential, 
shown in Ref. \cite{Camsonne:2013dfp} and reproduced here in Fig. \ref{ff}.

\end{document}